\documentclass[fleqn,usenatbib]{mnras}

\usepackage{newtxtext,newtxmath}

\usepackage[T1]{fontenc}

\DeclareRobustCommand{\VAN}[3]{#2}
\let\VANthebibliography\thebibliography
\def\thebibliography{\DeclareRobustCommand{\VAN}[3]{##3}\VANthebibliography}

\usepackage{graphicx}	
\usepackage{amsmath}	
\defcitealias{Chen_2022}{C22}
\usepackage{rotating}

\title[Radio-Loud Varstrometry Selected Quasars]{Varstrometry selected radio-loud candidates of dual and off-nucleus quasars at sub-kpc scales}

\author[Wang et al.]{
Hao-Chen Wang,$^{1,2}$\thanks{E-mail: hcw062@mail.ustc.edu.cn}
Jun-Xian Wang,$^{1,2}$\thanks{E-mail: jxw@ustc.edu.cn}
Min-Feng Gu$^{3}$
and Mai Liao$^{4,5,6}$
\\
$^{1}$CAS Key Laboratory for Research in Galaxies and Cosmology, Department of Astronomy, University of Science and Technology of China, Hefei, Anhui 230026, China\\
$^{2}$School of Astronomy and Space Science, University of Science and Technology of China, Hefei 230026, China\\
$^{3}$Key Laboratory for Research in Galaxies and Cosmology, Shanghai Astronomical Observatory, Chinese Academy of Sciences, Shanghai 200030, China\\
$^{4}$National Astronomical Observatories, Chinese Academy of Sciences, 20A Datun Road, Chaoyang District, Beijing 100101, China\\
$^{5}$Chinese Academy of Sciences South America Center for Astronomy, National Astronomical Observatories, CAS, Beijing, 100101, China\\
$^{6}$Instituto de Estudios Astrofísicos Facultad de Ingeniería y Ciencias Universidad Diego Portales Av. Ejército 441, Santiago, Chile
}

\date{Accepted XXX. Received YYY; in original form ZZZ}

\pubyear{2023}

\begin{document}
\label{firstpage}
\pagerange{\pageref{firstpage}--\pageref{lastpage}}
\maketitle

\begin{abstract}

Dual super massive black holes (SMBHs) at sub-kpc to kpc scales, the product of galaxy mergers, are progenitors of eventually coalescing binary SMBHs. If both or one of the dual SMBHs are accreting, they may appear as dual AGNs or off-nucleus AGNs. Studying such systems is essential to learn the dynamical evolution of binary SMBHs as well as the process of galaxy merging. Recently a novel astrometry-based method named varstrometry has been put forward to search for dual SMBHs at high redshift, as the unsynchronized flux variability of dual AGNs (or off-nucleus AGNs) will cause astrometric jitters detectable by \emph{Gaia} without spatially resolving them. Based on \emph{Gaia} varstrometry we select a rare sample of five radio loud quasars with clear \emph{Gaia} astrometric jitters. With e-MERLIN observations we have revealed a single compact radio source for each of them. Remarkably all but one exhibit clear \emph{Gaia}-radio offsets of $\sim$ 7--40 mas. 
The observed \emph{Gaia} jitters appear consistent with the expected values. 
These detected \emph{Gaia}-radio offsets suggest these candidate dual SMBHs may have projected separations as small as $\sim$0.01--0.1 arcsec ($\sim$0.1 kpc, depending on the optical flux ratio of two SMBHs). Meanwhile, this work highlights the remarkably high efficiency of \emph{Gaia} varstrometry selection of jittering sources.  

\end{abstract}




\section{Introduction} \label{sec:intro}

Dual super massive black holes (SMBH) may appear in the post-merger stage of galaxies, which could play a key role in galaxy formation and evolution \citep[e.g.][]{Jaffe_2003, hughes_2009}. As time goes by, dual SMBHs may get closer to each other due to dynamical friction, form binary SMBHs with a compact orbit $<$10 pc in its own potential, and finally coalesce \citep[e.g.][]{Blaes_2002, khan_2013}. Searching for close ($\sim$kpc or sub-kpc) dual SMBHs, the progenitors of coalescing SMBHs, 
thus are essential to probe their dynamical evolution as well as the process of galaxy merging \citep{DeRosa2019}, especially at high redshifts when galaxy mergers occurs much more frequently. Such dual SMBHs may appear as a dual AGN if both SMBHs are accreting, or an off-nucleus AGN, if only one SMBH is active.
A number of dual AGNs have been reported in literature \citep[e.g.][]{Woo_2014,Li_2016,Ellison_2017,Shen_2021,Chen_2022,Mannucci_2022,Koss_2023}, the majority of which (as shown in Fig. 1 of \citealt{Chen_2022}, hereafter \citetalias{Chen_2022}) however are at low redshifts and/or have separations $>$10 kpc.   

Along with {\it Gaia's} outstanding astrometric performance and full sky coverage, a novel astrometric method to search for $\sim$kpc scale dual SMBHs at high redshifts has been proposed \citep{Hwang_2020}, which is named as varstrometry.
Considering a pair of closely tied active SMBHs with small angular separation, the photocenter of the unresolved system is dependent of the relative brightness of each component. 
Owing to the unsynchronized stochastic flux variation of each AGN, the photocenter may change its location noticeably enough for {\it Gaia's} detection without spatially resolving the pair. For off-nucleus AGNs, the photocenter of the variable AGN emission and the stable host galaxy flux may change as well. The changing photocenter can be well quantified by the parameter \emph{astrometric\_excess\_noise} \citep{Lindegren2018}, calculated by {\it Gaia's} astrometric solution to describe the disagreement between observations and the best-fitting standard astrometric model.
Meanwhile, due to unequal optical path for each lensed quasar image, unresolved lensed quasars may result in the same observational effect (astrometric jittering) as dual and off-nucleus AGNs do. Similarly quasar-star superpositions in optical image may also yield astrometric jitters \citep{Shen_2021,Lemon_2017,Chen_2022}. 

\begin{figure}
    \centering
    \includegraphics[scale=0.435]{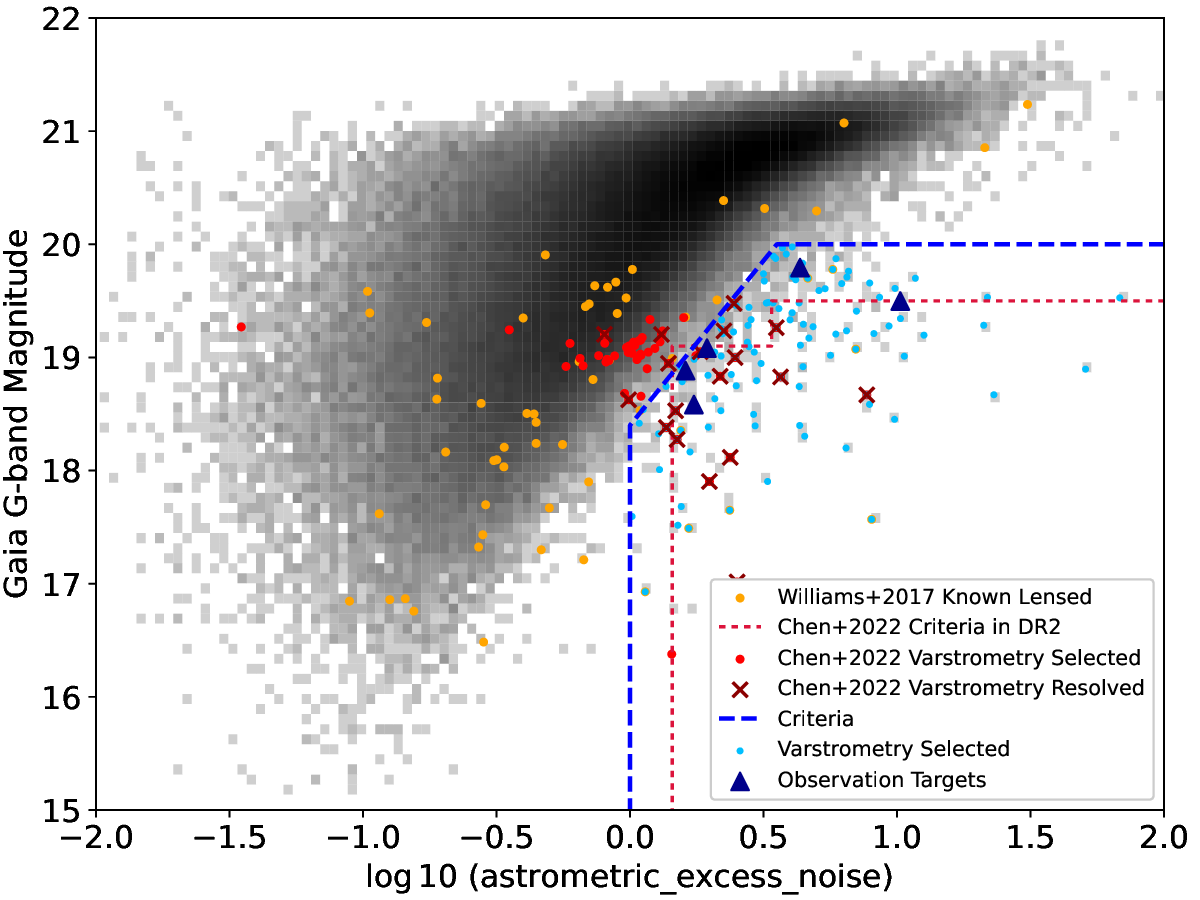} 
    \caption{\emph{Gaia} G-band magnitude versus \emph{astrometric\_excess\_noise} of SDSS quasars (grey scale image). Our varstrometry selection criteria for candidate dual and off-nucleus AGNs are shown as blue dashed lines. The $z$ $>$ 0.5 candidates are shown as blue dots, among them five with FIRST 1.4 GHz flux above 5 mJy are our selected targets for e-MERLIN observations (blue triangles). Known lensed quasars from \citealt{Williams_2017} at $z$ $>$ 0.5 are plotted as orange dots. The varstrometry selection criteria of \citetalias{Chen_2022} using \emph{Gaia} DR2 are drawn as red dashed lines, and their varstrometry selected quasars (but now plotted using \emph{astrometric\_excess\_noise} from \emph{Gaia} EDR3 instead of DR2) are marked as red dots with 18 HST resolved ones marked as red crosses.}
    \label{fig:select}
\end{figure}

Recently \citetalias{Chen_2022} presented HST snap images of 56 \emph{Gaia}-unresolved candidates dual quasars selected with varstrometry (based on \emph{Gaia} DR2, \citealt{gaia_2018}), 18 of which show two or multiple sub-arcsecond cores (with separations as small as 0.2 arcsec) in HST images, confirming the remarkable efficiency of the varstrometry approach.  
While a significant fraction of the resolved pairs could be physical quasar pairs or lensed quasars with separations of $\sim$several kpc \citepalias{Chen_2022}, the nature of those \emph{HST} unresolved candidates is unclear but also attractive. Are they reliable jittering sources? If yes, some of them are likely dual or off-nucleus AGNs with even smaller separations (sub-kpc), for which resolving the pair or detecting the small offset between the active SMBH and the host galaxy could be beyond the ability of current optical instruments. 

In this letter, we choose varstrometry selected candidate dual and off-nucleus quasars utilizing \emph{Gaia} EDR3 \citep{Gaia_EDR3_2021}, which provide significantly better astrometric measurements compared with earlier releases \citep{Lindegren2021}. We present radio (e-MERLIN) observations of five radio loud candidates. We show that, high resolution radio interferometric observations (of radio loud candidates) can robustly confirm the remarkable efficiency of the varstrometry selection, and dramatically help us to dive deeper into the nature of the selected candidates. In \S\ref{sec:obs}, we present our candidates selected from \emph{Gaia} EDR3 and SDSS DR16Q. We obtain e-MERLIN observations on five candidates, with the analyses and results given in  \S\ref{sec:results}, followed by a discussion on \S\ref{sec:dis}. Through out this work, we adopt a flat Lambda cold dark matter cosomology with $\Omega_m=0.31$, $\Omega_\Lambda=0.69$, $H_0=$ 67.7 km $\rm{s}^{-1}$ $\rm{Mpc}^{-1}$ \citep{Planck2018_2020}.

\section{Target Selection} \label{sec:obs}

Here we first describe our procedures to select candidate dual and off-nucleus AGNs based on \emph{Gaia} varstrometry. Firstly, we perform a cross match between \emph{Gaia} EDR3 \citep{Gaia_2016,Gaia_EDR3_2021} and SDSS DR16Q \citep{SDSS_DR16Q_2020} with a radius of 3 arcsec, and pick the closest match when multiple \emph{Gaia} EDR3 matches are found. It has been proposed that \emph{astrometric\_excess\_noise} $>$ 1 is a good criterion to select jittering sources \citep{Lindegren2018,Hwang_2020}. However, in Fig. \ref{fig:select} we show that \emph{astrometric\_excess\_noise} shows clear dependence on \emph{Gaia} magnitude, indicating a dominant fraction of the large \emph{astrometric\_excess\_noise} values ($>$1) of faint sources are likely artificial \footnote{As noted in \cite{Hwang_2020}, the reliability of \emph{astrometric\_excess\_noise\_sig} (an indicator of whether \emph{astrometric\_excess\_noise} is statistically significant) however is yet unclear. We would like to defer the discussion and use of \emph{astrometric\_excess\_noise\_sig} to a future work.}.
We optimize the criteria to select candidates: 1) $z$ $>$ 0.5, to eliminate the strong effect of extended host emission on the accuracy of astrometric variability \citep{Hwang_2020}; 2) a magnitude-dependent threshold of  \emph{astrometric\_excess\_noise} (see Fig. \ref{fig:select}); 3) \emph{Gaia} {\it G}-band magnitude brighter than 20. 111 quasars satisfy our criteria. 

\begin{table*}
	\centering
	\caption{Five candidates of off-nucleus or dual quasars with e-MERLIN observations}
	\label{tab:obs_info}
    \resizebox{\linewidth}{!}{
	\begin{tabular}{ccp{0.2\columnwidth}p{0.2\columnwidth}ccccccc} 
		\hline
		Name & Redshift & \emph{Gaia} position & Radio position &  Separation & \emph{Gaia} RA error & \emph{Gaia} dec. error & Radio RA error & Radio dec. error & \emph{astrometric\_excess\_noise} & Core flux density\\
         &  & \quad\enspace(ICRS) & \quad\enspace(ICRS) & [mas] & [mas] & [mas] & [mas] & [mas] & [mas] & mJy/beam\\
		\hline
		J1044+2959 & 2.98 & 10:44:06.34263 +29:59:01.0030 & 10:44:06.34321 +29:59:00.9959 & 10.4 & 0.244 & 0.265 & 1.60 & 0.45 & 1.614 & 137.6\\
        J1325+0412 & 0.73 & 13:25:52.63408 +04:12:00.7777 & 13:25:52.63118 +04:12:00.7738 & 43.5 & 0.752 & 0.425 & 3.25 & 4.38 & 4.333 & 0.43\\
        J1433+4842 & 1.36 & 14:33:33.03160 +48:42:27.7752 & 14:33:33.03113 +48:42:27.7811 & 7.5 & 0.207 & 0.227 & 3.18 & 2.73 & 1.942 & 0.95\\
        J1733+5520 & 1.20 & 17:33:30.84382 +55:20:30.8526 & 17:33:30.84310 +55:20:30.8476 & 7.9 & 0.198 & 0.218 & 1.07 & 0.94 & 1.735 & 6.26\\
        J2109+0656 & 2.94 & 21:09:47.09470 +06:56:34.7520 & 21:09:47.09314 +06:56:34.7539 & 23.4 & 1.173 & 1.043 & 1.32 & 4.38 & 10.287 & 3.19\\
		\hline
	\end{tabular}}
\end{table*}

\begin{figure*}
    \centering
    \includegraphics[scale=0.7]{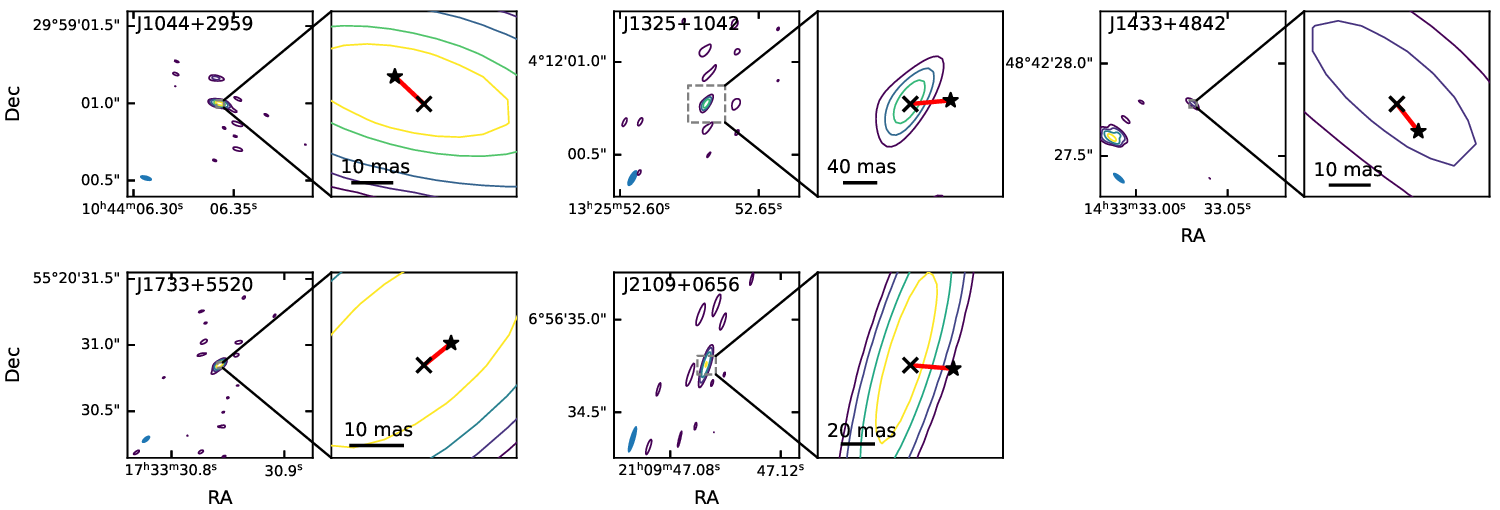}
    \caption{e-MERLIN images of the five sources. The left panel for each source shows the radio images with the primary beam shown on the bottom left. The contour levels are (3,6,12,24, and 32) $\times$ 3.05 mJy/beam (rms, the same below) for J1044+2959, (3,5,7, and 9) $\times$ 0.04 mJy/beam for J1325+1042, (2.5,5,10, and 20) $\times$ 0.14 mJy/beam for J1433+4342, (3,6,12, and 24) $\times$ 0.16 mJy/beam for J1733+5520 and (3,6,12, and 18)
    $\times$ 0.12 mJy/beam for J2109+0656. The right-hand panel shows the zoom-in radio image centring at each sources. The positions of compact radio cores are drawn as crosses and the positions of \emph{Gaia} optical centres are drawn as stars, which are connected with red lines. Significant \emph{Gaia}-radio offsets are revealed in all sources but J1433+4842.}
    \label{fig:obs}
\end{figure*}

In Fig. \ref{fig:select} we over-plot the 56 varstrometry selected quasars of \citetalias{Chen_2022} with followup \emph{HST} observations, 18 of which are spatially resolved into dual or multiple cores by \emph{HST}. Our criteria could include 16 of the candidates of \citetalias{Chen_2022}, 14 out of which have been resolved by \emph{HST} images. 
This indicates while our criteria is highly effective in selecting jittering sources (likely with a success rate of 14/16 = 87.5 per cent), 
the current selection region could be expanded in the future to include more candidates. 
Meanwhile, although the selection criteria of \citetalias{Chen_2022} (see Fig. \ref{fig:select}) is more conservative than ours, a significant portion of their candidates show rather low \emph{astrometric\_excess\_noise} in \emph{Gaia} EDR3. This is because the selection of \citetalias{Chen_2022} was based on \emph{Gaia} DR2 while EDR3 we utilized provides significantly better astrometric measurements thus more reliable \emph{astrometric\_excess\_noise} \citep{Lindegren2021}. As an illustration, using the same selection criteria we adopted in this work but \emph{Gaia} DR2 data, we would select 5147 candidates (instead of 111). 

Only 7 candidates out of the 111 selected ones have FIRST 1.4 GHz radio flux larger than 5 mJy. The cut is applied to select radio bright sources for follow up high-resolution observations to accurately locate their radio core(s). 
Among them, J0749+2255 has been confirmed as double quasars by \cite{Shen_2021} with \emph{HST} and VLBA images (also included in \citetalias{Chen_2022}); and J1415+1129 has been identified as a lensed quasar \citep{Chartas_2004}. 
We selected the rest five sources as our e-MERLIN observation targets (see Table \ref{tab:obs_info}). Assuming a radio spectra index of 0.5, their radio loudness 
$R = f_{\rm 5GHz}/f_{\rm 2500A}$ range between 33 and 197. Note one of the targets, J1325+0412, is a type 2 AGN with a spectral redshift of 0.73.

\section{e-MERLIN Observations} \label{sec:results}

We observed the 5 sources with e-MERLIN in \emph{C}-band between 2021.8.15 and 2021.9.6 (CY12207). MK2, Kn, De, Pi and Cm were used for observations with three spectral windows of 128 MHz centering at 5 GHz. The observations were performed in phase-reference mode and lasted for $\sim$6 hours on each source. Radio images were calibrated with standard e-MERLIN pipeline.
We cleaned the radio images using \emph{tclean} task in {\small CASA} (\citealt{casa_2022}) with multi frequency synthesis and Briggs weighting. We set the robust parameter to 0.5 to reach a balance between spatial resolution and sensitivity. The clean beam size is 0.09 arcsec $\times$ 0.03 arcsec in average. The radio images of five sources are shown in Fig. \ref{fig:obs}.

We detect one compact radio core close to the optical position for each target. Fitting the compact radio core in each radio image with a two-dimensional Gaussian function using \emph{imfit} task in {\small CASA}, we obtain the positions and positional uncertainties\footnote{All the phase calibrators we used have position uncertainties less than 0.2 mas thus their uncertainties are ignored, and the stochastic contribution from atmospheric phase fluctuations are also negligible.} of radio core (see Table. \ref{tab:obs_info}). 
We use normalized separations \citep{Mignard_2016, EDR3_CRF3_2022} to evaluate the significance of \emph{Gaia}-radio offsets, taking account of the positional uncertainties of RA and Dec. and their covariance, for both \emph{Gaia} and radio coordinates. The covariance between RA and Dec., uncertainties of the compact radio core is calculated using the position angle of the best-fitting two-dimensional Gaussian from \emph{imfit}. All but one of five candidates (J1433+4842) have normalized separations $>$ 20 , reaching a confidence level of $1 - 10^{-86}$. 

On the other hand, for J1433, considering the 1$\sigma$ confidence range of the best-fitting positional angle, its normalized separation could be as low as 2.19 (corresponding to a confidence level of 90.9 per cent).
It indicates that the \emph{Gaia}-radio offset of J1433+4842 is relatively small and statistically insignificant. We found that, there is an extra source in \emph{Gaia} EDR3 near J1433+4842 within 1.09 arcsec.
As \citealt{Mannucci_2022} pointed out, the elongated photometric window on \emph{Gaia} allowed pair of sources with separations near or closer than 1.1 arcsec to be recorded as a single entry, possibly causing strange astrometric solutions and enlarging the \emph{astrometric\_excess\_noise}. The non-zero \emph{ipd\_frac\_multi\_peak} of J1433+4842 supports this scenario as well.
Therefore, although can not be ruled out (see further discussion in \S\ref{sec:dis}), the observed jittering of J1433+4842 is possibly artificial due to the contamination of the nearby source.
Note for each of the rest 4 targets, there is only a single \emph{Gaia} match within 3 arcsec.

\begin{figure}
    \centering
    \includegraphics[scale=0.4]{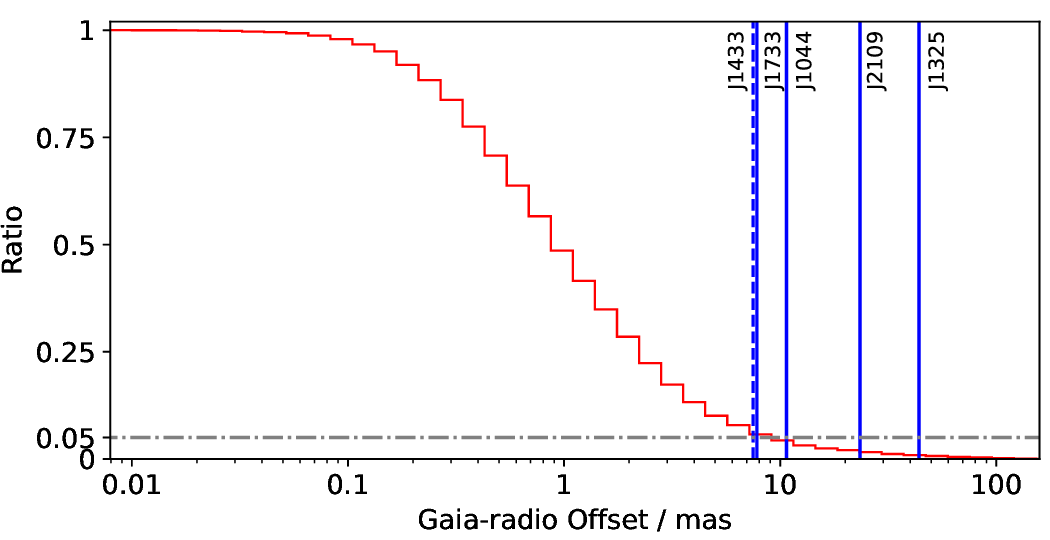}
    \includegraphics[scale=0.408]{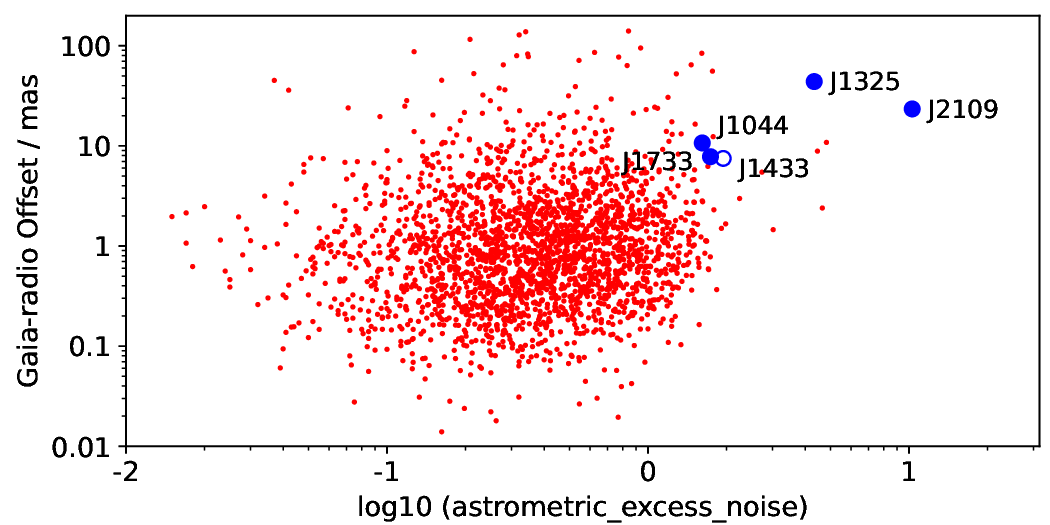}
    \caption{
    Top: \emph{Gaia}-radio offsets of our five candidates of dual and off-nucleus quasars, and the reverse-cumulative histogram of a control sample (red line). The control sample is derived through cross-matching the VLBI-based Radio Fundamental Catalogue (RFC) rfc\_2022d with \emph{Gaia} EDR3 and the Million Quasars catalog \citep{Flesch_2021}, with only sources with $z$ $<$ 0.5 and \emph{Gaia} \emph{G}-band magnitude $<$ 20 selected (similar with our targets). See http://astrogeo.org/sol/rfc/rfc\_2022d/ for rfc catalog.
    Bottom: \emph{Gaia}-radio offsets - \emph{astrometric\_excess\_noise} diagram of our candidates and the control sample.
    }
    \label{fig:offset}
\end{figure}

As shown in the upper panel of Fig. \ref{fig:offset},
all but one of our five candidates show significant and large \emph{Gaia}-radio offsets (as large as 44 mas). 
This is highly remarkable as the comparison between VLBI and \emph{Gaia} astrometry of a control sample of quasars (see Fig. \ref{fig:offset}) revealed $>10$ mas offsets only in 3.7 per cent sources. 
The lower panel of Fig. \ref{fig:offset} also shows that most quasar in the control sample have small \emph{astrometric\_excess\_noise} (only 0.4 per cent have \emph{astrometric\_excess\_noise} $>$ 1.614, which is the minimum value of our five targets).
Clearly, the observed large \emph{Gaia}-radio offsets of our candidates are associated with their large \emph{astrometric\_excess\_noise} values. 

The large \emph{Gaia}-radio position offsets and large \emph{astrometric\_excess\_noise} of our small sample indicate we have obtained a distinct and rare population of quasars with clear \emph{Gaia} astrometry jitters, and with the radio-loud AGN offset from the optical centre of the system, as expected in dual AGNs or off-nucleus AGNs. 

\section{Discussion} \label{sec:dis}
Suppose the mean optical flux and its variation (standard deviation) for each quasar in dual quasars, or for the off-nucleus quasar and its host galaxy, are $f_1$, $\sigma_1$ and $f_2$, $\sigma_2$, respectively ($\sigma_2$ = 0 for the host galaxy for off-nucleus quasar), and the projected distance between two components is $D$. The total optical flux and its standard deviation of the system is $f=f_1+f_2$ and $\sigma=\sqrt{\sigma_1^2+\sigma_2^2}$.
The positioning uncertainty of the optical center could be calculated as following, corresponding to \emph{astrometric\_excess\_noise} \citep{Hwang_2020}.

\begin{equation}
    \sigma_c=\frac{D}{(f_1+f_2)^2}\sqrt{\sigma_1^2f_2^2+\sigma_2^2f_1^2}
\end{equation}

The \emph{Gaia}-radio offset is the projected distance between the radio core (source 1) and optical center of the system:

\begin{equation}
    d_{GR}=\frac{f_2}{f_1+f_2}D
\end{equation}

Combining these equations, we get the predicted \emph{astrometric\_excess\_noise} as:

\begin{equation}
    \sigma_c=\frac{d_{GR}}{f}\sqrt{\sigma_1^2+\sigma_2^2f_1^2/f_2^2}
\end{equation}

\begin{figure}
    \centering
    \includegraphics[scale=0.4]{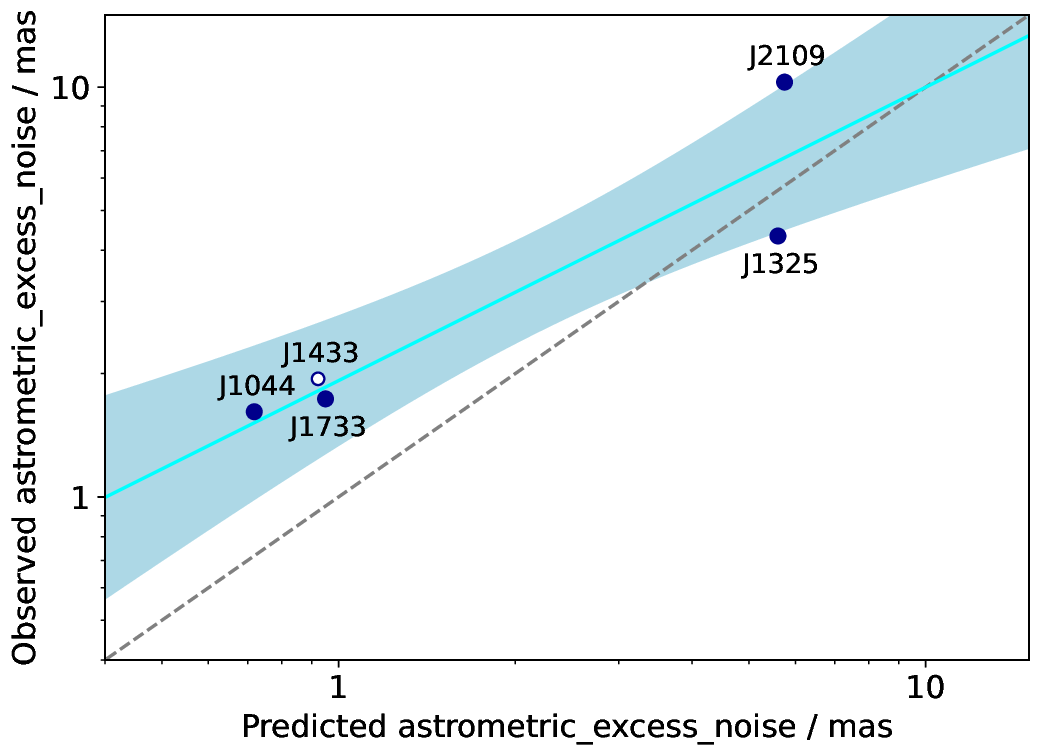}
    \caption{Observed and predicted \emph{astrometric\_excess\_noise} for each candidates assuming they are all off-nucleus quasars. The blue line is the linear regression of the sample (excluding J1433+4842) with the 1$\sigma$ area of the regression filled with light blue.}
    \label{fig:predicted}
\end{figure}

Assuming that our candidates are off-nucleus quasars, or dual quasars but optical flux variation in one of them is negligible, i.e., $\sigma_2$ = 0, or dual quasars with $f_1$ $\sim$ $f_2$ (assuming the host galaxy contribution is negligible), we may calculate the predicted \emph{astrometic\_excess\_noise} as 

\begin{equation}
    \sigma_c=d_{GR}\frac{\sigma}{f}
\end{equation}
where $\sigma$ is the flux variation\footnote{ For each \emph{Gaia} source, its observed flux variation could be derived as 
\emph{phot\_g\_mean\_flux\_error} * $\sqrt{\mbox{\emph{phot\_g\_n\_obs}}}$, where \emph{phot\_g\_mean\_flux\_error} is the mean \emph{Gaia} \emph{G}-band flux uncertainty and \emph{phot\_g\_n\_obs} the number of \emph{Gaia} \emph{G}-band exposures, both given in \emph{Gaia} EDR3. 
The intrinsic flux variation $\sigma$ is then obtained through subtracting in quadrature the expected \emph{Gaia} photometric uncertainty per exposure. 
The expected \emph{Gaia} photometric uncertainty per exposure as a function of magnitude is calculated using the tool provided by \emph{Gaia} DPAC to reproduce the \emph{Gaia} (E)DR3 photometric uncertainties described in the GAIA-C5-TN-UB-JMC-031 technical note using data in \cite{Riello2021}. See https://www.cosmos.esa.int/web/gaia/fitted-dr3-photometric-uncertainties-tool.  For all of our five quasars, the photometric uncertainties only make small contribution to the observed variation.} of our candidates,
and compared it with observed value (see Fig. \ref{fig:predicted}). 

A correlation (with coefficient $R$ of 0.917 and p-value of 0.140)
is found between the observed and predicted \emph{astrometric\_excess\_noise}, and the best-fitting
linear regression line is close to 1:1. This is similar to Fig. 6 of \cite{Hwang_2020}, who calculated the expected \emph{astrometric\_excess\_noise} for a small sample of \emph{Gaia} unresolved pre-main sequence binaries (but resolved by other facilities, thus with the measurements of $D$ instead of $d_{GR}$).

Our Fig. \ref{fig:predicted} thus provides new and remarkable support to the varstrometry approach, that the observed \emph{astrometric\_excess\_noise} provides a reasonable measurement of the astrometric jittering in quasars. 
We note that, though the observed \emph{Gaia}-radio offset in J1433+4842 is statistically insignificant, its observed \emph{astrometric\_excess\_noise} also appears consistent with the expected value, suggesting its jittering could be physical real. Therefore we keep this source as a possible candidate of dual or off-nucleus quasars, and await future deep radio images to confirm its \emph{Gaia}-radio offset.

Furthermore, we may calculate the projected distance between two quasars (or between the off-nucleus quasar and its host galaxy center) in our candidate systems assuming hypothetical brightness ratio $q=f_1/f_2$
ranging from 1:10 to 10:1 (assuming quasar 1 is the radio detected one):

\begin{equation}
    D = \frac{f_1+f_2}{f_2}d_{GR} = (1+q)d_{GR}
\end{equation}
Note the brightness ratio of the \emph{HST} resolved pairs in \citetalias{Chen_2022} is between 1.0 and 8.7 with a median of 2.1 (the flux ratio of the brighter source to the fainter one). 

\begin{figure}
    \centering
    \includegraphics[scale=0.38]{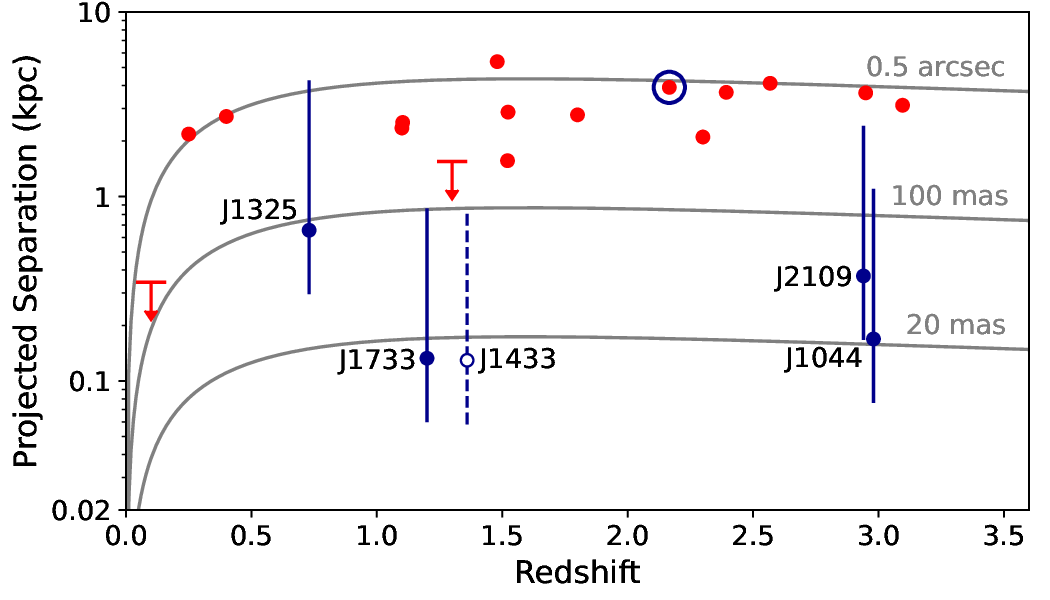}
    \caption{The expected projected separation of our five candidate dual quasars, in comparison with the \emph{HST} measured projected separation of the varstrometry selected quasars of \citetalias{Chen_2022} which also satisfy our selection criteria based on \emph{Gaia} EDR3 (an upper limit of 0.18 arcsec, the smallest resolved separation in \citetalias{Chen_2022}, is assigned to the two HST unresolved ones).
    For our targets, the expected separation is calculated from the measured \emph{Gaia}-radio offset assuming the optical flux ratio of the two quasars (of the e-MERLIN detected quasar to the other) of 1:10, 1:1 and 10:1 (the lower limits of the blue error bars, blue dots, and upper limits, respectively). 
    One of two other varstrometry selected radio quasars, J0749+2255, which is also included in  \citetalias{Chen_2022}, in marked as an open blue circle.  The other one, 
     J1415+1129, a lensed quasar with four images spacing in $\sim$1 arcsec, is omitted in this figure. 
     }
    \label{fig:sep}
\end{figure}

The results are presented in Fig. \ref{fig:sep}. We over-plot the 16 varstrometry selected quasars of \citetalias{Chen_2022} which fall in our selection region (see Fig. \ref{fig:select}), 14 of them are resolved with \emph{HST}. 

Compared to the sample of \citetalias{Chen_2022}, though to be statistically confirmed with larger sample, our sample tends to have considerably smaller projected distances. One possibility is due to the difference in the selection between \citetalias{Chen_2022} and this work, and in the measurements of \emph{astrometric\_excess\_noise} between \emph{Gaia} DR2 and EDR3. 
As shown in Fig. \ref{fig:select}, the selection of \citetalias{Chen_2022} appears more conservative, thus favors targets with larger \emph{astrometric\_excess\_noise} (thus larger projected distance). 
Meanwhile, \emph{Gaia} DR2 provides less reliable measurements of small \emph{astrometric\_excess\_noise}, thus is less efficient in selecting sources with small projected distances. 
Since we have demonstrated the remarkably high efficiency of varstrometry selection of jittering sources based on \emph{Gaia} EDR3, \emph{HST} followup observations on a sample uniformly selected based on \emph{Gaia} EDR3 or later releases are thus essential to statistically study the population, and \emph{HST} unresolved candidates among them could consequently be considered as reliable jittering sources, thus potential dual and off-nucleus AGNs at smaller scales.

Alternatively, varstrometry selected radio loud quasars may intrinsically be different. 
Through matching \emph{Gaia} DR2 with the VLBI-based Radio Fundamental
Catalog, \cite{Plavin_2018} found radio AGNs with clear \emph{Gaia}-VLBI offsets often have the direction of offset aligning with the jet, indicating strong optical jet contamination to \emph{Gaia} astrometry could cause \emph{Gaia}-radio offset and possibly also jittering \citep[see also][]{Kovalev2017,Petrov2017}. 
As shown in the lower panel of Fig. \ref{fig:offset}, while our candidates have both large \emph{Gaia}-radio offsets and \emph{astrometric\_excess\_noise}, most of the VLBI control sample have rather small offsets and \emph{astrometric\_excess\_noise}.  
We further find, that \emph{Gaia}-radio offsets of the control sample barely change with increasing \emph{astrometric\_excess\_noise}.
This suggests our small sample is physically distinct from the control sample, and the physical origin of their jitter and \emph{Gaia}-radio offset is unlikely due to jet contamination.
Meanwhile, except for J1044+2959, our sources have rather low radio core fluxes (0.4--6 mJy), considerably fainter than the VLBI \emph{C}-band fluxes (with a median of 69 mJy) of the control sample at similar \emph{Gaia} \emph{G} magnitude.  
Furthermore, we only detected compact radio cores (but not extended jet like structure) in e-MERLIN images of our five targets. This suggests our sources do not have jet as strong as RFC sources, thus the optical jet contribution to \emph{Gaia} astrometry of them is expected to be weak. Future deep radio images could put further constraints to the existence of strong jets in our targets.

A more interesting possibility is that varstrometry selected radio loud quasars do intrinsically have smaller projected distances. 
If future deep radio images (e.g. SKA) could confirm that radio loud varstrometry selected dual and off-nucleus quasars preferentially have smaller projected distances (sub-kpc versus kpc) compared with radio quiet ones, 
this could interestingly imply that dual SMBHs or offset SMBH at sub-kpc scales are more likely to trigger jet launching compared with those at kpc scales. 

A considerable fraction of varstrometry selected quasars could be quasar-star pairs or lensed quasars \citep[e.g.][]{Shen_2021, Chen_2022}. 
As Fig. \ref{fig:sep} shows, our small sample seems to have shorter projected separations ($\lesssim$0.1 arcsec) compared to the "dual/lensed quasars" in \citetalias{Chen_2022} ($\sim$0.5 arcsec), resulting in 25 times lower possibility of finding close quasar-star pairs. Among the 16 targets of \citetalias{Chen_2022} which satisfy our selection criteria, nine are identified as "quasar-star" pairs, thus for our small sample with much smaller predicted separations, the fraction of quasar-star pairs is expected to much lower (9/16/25 = 2.25 per cent).

On the other hand, while lensed quasars can not be completely ruled out, our high resolution e-MERLIN observations revealed no evidence of lensed signals (arcs, rings, or multiple images), though this possibility could be further testified with future deep radio images.  
Furthermore, \cite{Shen_2021} stated that the abundance of high-redshift sub-arcsecond gravitational lens is insufficient to account for their varstrometry selected resolved pairs. 

It is also interesting to note that numerical simulations \citep[e.g.][]{Blecha2018} have suggested that AGNs in the sub-kpc merger stage should be heavily obscured. Varstrometry selected sub-kpc dual AGNs seems contradict this scenario, or we are selecting a sub-population of them in which at least one AGN is unobscured. Further note one of our candidate J1325+0412 is indeed a type 2 AGN. If it is in a dual AGN system, its jitter and flux variability measured with \emph{Gaia} shall be caused by the other unobscured AGN in the system (which however was not seen in SDSS spectrum), unless itself changed to type 1 during \emph{Gaia} observations. Followup spectroscopic observation is required to verify if this sources is a changing-look quasar \citep[e.g.][]{LaMassa2015} in dual or off-nucleus AGN system.

Besides \emph{astrometric\_excess\_noise}, other criteria may also reveal candidate dual quasars out of \emph{Gaia} unresolved sources. As \cite{Gaia_EDR3_ruwe} presented, renormalised unit weight error (\emph{ruwe}), which describes the goodness of fit with standard astrometric model similar to reduced $\chi^2$, could be utilized to select candidate dual quasars (normal stars and quasars with good astrometric behaviour should have \emph{ruwe} close to 1, and sources with \emph{ruwe} $>$ 1.4 are possibly dual quasar candidates). Four of our five e-MERLIN target fullfilled this criteria except J1044+2959 (whose \emph{ruwe} is 1.34, also close to 1.4). 

Additionally, \cite{Mannucci_2022} proposed the \emph{Gaia} Multi-Peak method and successfully found dual/lensed AGN candidates at sub-arcsec separations by looking for multiple peaks in the light profiles observed by \emph{Gaia}. As mentioned in \S\ref{sec:obs}, close pairs with separations $\leq$1.1 arcsec may be observed and recorded by single elongated photometric window of \emph{Gaia}, whose size is as large as 0.71 arcsec $\times$ 2.1 arcsec. In this case, multiple peaks can be seen in the compressed one-dimensional light profile, and \emph{ipd\_frac\_multi\_peak} is the fraction of \emph{Gaia} observations with multiple peaks detected. A quasar with large \emph{ipd\_frac\_multi\_peak} is likely to be a dual/lensed quasar. 
Among our five targets, only J1733+5520 shows a large \emph{ipd\_frac\_multi\_peak} of 20 per cent. As J1733+5520 is unresolved by \emph{Gaia}, its possible separation is likely between 0.11 arcsec (\emph{Gaia}'s PSF) and 0.35 arcsec \citep{Mannucci_2022}, roughly consistent with the range of 0.009-0.09 arcsec we derived based on its \emph{Gaia}-radio offset.

Moreover, \cite{Wu_2022} selected abnormal \emph{Gaia} unresolved quasars (e.g. quasar pairs). 
\cite{Makarov_2022} yielded a sample of 152 quasars with excess proper motions, which were candidates of dual/multiple AGNs, or lensed quasars. 
Note two of our candidates (J1733+5520 and J2109+0656) do exhibit highly significant \emph{Gaia} DR3 proper motions, with covariance-normalized values of 16.0 and 11.8 respectively, and J2109+0656 has an uncertainty-normalized parallax of +4.3.
Such proper motion and parallax should be artificial due to the jittering of the sources. For instance, a monotonic flux increasing of one quasar in dual AGN system could cause artificial proper motion signal. 
Further crossmatch studies of the parent sample in this letter (111 varstrometry selected quasars) with those samples aforementioned, and extensively exploring the efficiencies of these various selection criteria is deferred to a future work, better with later \emph{Gaia} data releases when epoch astrometry is available.

\section*{Acknowledgements}
We thank the anonymous referee for constructive comments which have significantly improved the manuscript.
This work was supported by the National Science Foundation of China (No. 1890693, 12033006 $\&$ 12192221), and Cyrus Chun Ying Tang Foundations.
MFG acknowledges support from the National Science Foundation of China (grant no. 11873073), Shanghai Pilot Program for Basic Research Chinese Academy of Science, Shanghai Branch (JCYJ-SHFY2021-013), the National SKA Program of China (Grant No. 2022SKA0120102), the Original Innovation Program of the Chinese Academy of Sciences (E085021002), and the science research grants from the China Manned Space Project with No. CMSCSST-2021-A06.
e-MERLIN is a National Facility operated by the University of Manchester at Jodrell Bank Observatory on behalf of STFC, part of UK Research and Innovation.
This work has made use of data from the European Space Agency (ESA) mission
\emph{Gaia} (\url{https://www.cosmos.esa.int/gaia}), processed by the \emph{Gaia}
Data Processing and Analysis Consortium (DPAC,
\url{https://www.cosmos.esa.int/web/gaia/dpac/consortium}). Funding for the DPAC
has been provided by national institutions, in particular the institutions
participating in the \emph{Gaia} Multilateral Agreement.\\

\section*{Data Availability}
The data underlying this article will be shared on reasonable request to the corresponding author.




\bibliographystyle{mnras}
\bibliography{paper_final} 








\bsp	
\label{lastpage}
\end{document}